# Size effect on deformation twinning at the nanoscale in hcp metals

Ya-Fang Guo,[*,1] Shuang Xu,[1] Xiao-Zhi Tang,[1] Yue-Sheng Wang,[1] and Sidney Yip[2]

[1] Institute of Engineering Mechanics, Beijing Jiaotong University, Beijing 100044, China
[2] Department of Nuclear Science and Engineering, Massachusetts Institute of Technology, 77 Massachusetts Avenue, Cambridge, Massachusetts 02139, USA

**ABSTRACT:** Deformation twinning is generally considered to be the primary mechanism for hexagonal close-packed (hcp) metals due to their limited slip systems. Recent microcompression experiments on hcp metals point to intriguing strong size effects on twinning mechanisms, indicating that pyramidal slips dominate compression.[1-4] In this work we analyze size effects on twinning due to lattice-rotation strain in hcp single crystals. A criterion for deformation twinning is derived at the nanoscale and tested by molecular dynamics simulations of magnesium and titanium single crystals. The results show <c+a> pyramidal slip dominates the compression deformation at the nanoscale, consistent with experimental observations [1-4] as well as our analysis. Our finding clarifies the nature of size effects in deformation twinning, while also providing an explanation for the so-called strength differential (SD) effect.

Because the number of slip systems in hexagonal close-packed (hcp) metals is limited, deformation twinning can be favored over dislocation slip. In magnesium and its alloys, two types of c-axis twins are frequently reported, $\{10\bar{1}2\}\langle10\bar{1}1\rangle$ tension [5-7] and $\{10\bar{1}1\}\langle10\bar{1}2\rangle$ compression.[8-10] Generally, compression twins are thought to be the primary mechanism for c-axis compression on bulk samples.[8,9] However, in recent years microcompression experiments on hcp metals [1-3] have revealed intriguing aspect of twinning mechanism at the nanoscale, namely, a pronounced size effect on deformation in compression. From in situ nano-compression measurements in hcp titanium, it was found for single crystal samples below 1 μm in size, deformation twinning was entirely replaced by dislocation plasticity.[1] At the same time twinning was not observed in sample sizes from 2.1 to 10 μm under c-axis compression in microcompression measurement of single crystal FIB columns on (0001) magnesium.[2] In this case significant plasticity and hardening occurred due to six active pyramidal π2 slip systems, and there was appreciable size effect in the flow strength, the smaller being stronger. Additionally TEM analysis of micropillar specimens of magnesium single crystals showed that no deformation twinning in compression along the [0001] c-axis.[3] Multiple slip systems become active on the pyramidal planes, resulting in significant hardening for length scales from 2.5 to 10 µm.

Besides experimental observations, there are also indications from molecular dynamics (MD) simulations of deformation in magnesium single crystal showing no deformation twinning under compression, but dominance of pyramidal <a+c> slip.[11] Furthermore, in simulation of c-axis tension both tension and compression twinning have been observed.[12] These simulation results along with the experimental observations therefore suggest the prevailing notion that compression twin would be activated when there is a contraction strain component parallel to the c-axis [8,9] may be incomplete. That size effects associated with the hcp lattice structure are relevant has been discussed recently.[11] A condition was derived for the onset of deformation twinning in compression which requires the c/a-ratio to be less than $\sqrt{3}$.

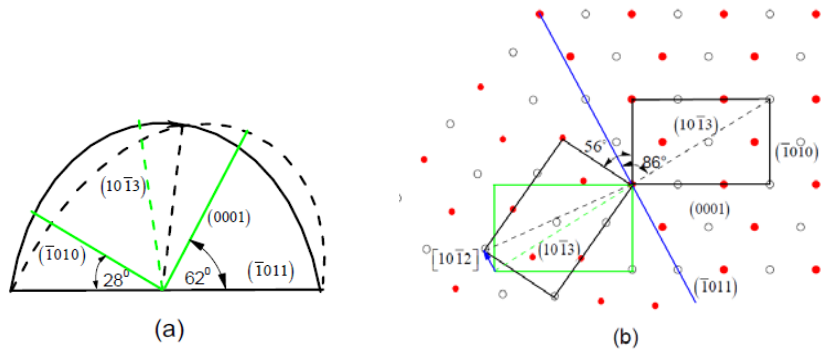

**Figure 1.** (a) Strain ellipsoid for magnesium revealing twin-related foreshortening of prism plane and extension of basal plane. (b) Atom arrangement of the $\{10\bar{1}1\}$ twin. The $K_1$, $K_2$ and $\varsigma_1$ are separately $\{10\bar{1}1\}$, $\{10\bar{1}3\}$ and $\langle 10\bar{1}2 \rangle$ for the $\{10\bar{1}1\}$ twinning. Shear occurs along the $[10\bar{1}2]$ direction for the $(10\bar{1}3)$ plane.

Previous analysis of twinning deformation is generally based on calculating the strain caused by the twinning shear.[13,14] According to the strain ellipsoid for hcp metals twinning by $\{10\bar{1}2\}$ mode cannot occur when $c/a = \sqrt{3}$. For $c/a < \sqrt{3}$, twinning will occur if compressive loads are applied parallel to the basal plane or tensile loads applied parallel to the prism planes. The opposite holds for $c/a > \sqrt{3}$, where twinning occurs when the prism plane is compressed or the basal plane extended. [13] For the $\{10\bar{1}1\}$ twin in magnesium, the $K_1$, $K_2$ and $\varsigma_1$ are $\{10\bar{1}1\}$, $\{10\bar{1}3\}$ and $\langle 10\bar{1}2 \rangle$, respectively. Shear occurs along the $[10\bar{1}2]$ direction for the $(10\bar{1}3)$ plane. According to the strain ellipsoid shown in Figure 1a, the $\{10\bar{1}1\}$ twinning will occur when compression is applied parallel to prism plane or tension applied parallel to basal plane. It was predicted that $\{10\bar{1}1\}$ twinning is favored under c-axis compression when c/a>1.5 if one considers only the strain caused by twinning shear. When c/a<1.5, $\{10\bar{1}1\}$ twinning will be favored under the c-axis tension [13]. Thus for general hcp metals such as magnesium and titanium (c/a>1.5), the $\{10\bar{1}1\}$ twin is a compression twin when the twinning shear is considered.

It appears twinning mechanisms for hcp metals are more complicated than what the foregoing considerations would lead us to believe. For example, an atomic shuffling dominated mechanism was reported for the $\{10\bar{1}2\}$ twinning in magnesium, suggesting the process involved the conversion of basal planes to prism planes.[15] For the $\{10\bar{1}1\}$ twin, atomic displacements that can be described as shear plus a series of shuffles have been revealed during twinning.[16] Thus $\{10\bar{1}1\}$ twinning cannot be activated by $\langle 10\bar{1}2 \rangle$ shear alone. Some complicated atomic movements, such as atoms shuffling, should be considered. Figure 1b shows the atomic arrangement of the $(\bar{1}011)$ twin. The original and twinned crystal structures are denoted by the green and black squares respectively. It can be seen one cannot be transformed into the other by $\langle 10\bar{1}2 \rangle$ shearing. To fully describe twinning in hcp metals, it is therefore not enough to consider only the strain caused by the twinning shear. A new criterion is needed that will take into account other factors that also affect twinning at the nanoscale.

The stacking-fault energy (SFE) is traditionally regarded as an indicator of tendency towards twinning in fcc metals, with materials having low SFE being favored.[17-20] Besides SFE, elastic strain energy, interface energy, and energy dissipated during deformation also may play significant roles. When deformation twinning occurs, a region of the crystal is transformed by the external loading into its mirror counterpart. Crystal rotation occurs with no volume change after twinning, while the twinned region is symmetrical to the original matrix about the twin plane. In a polycrystalline material or a bulk single crystal, the volume strain could be ignored because there is no volume change after twinning. In this case, the SFE is more relevant for favoring twinning because it is closely related to the nucleation and growth of a twin. On the other hand, for a single crystal at the nanoscale, rotation of the lattice due to twinning will cause the elongation or contraction of the crystal along the loading direction, thus inducing a change of the elastic strain of the system. Due to the small size and the single orientation of the crystal at the nanoscale, the elastic strain caused by lattice rotation cannot be cancelled out by other deformation. Under this condition, the elastic strain energy may play a more important role than the stacking-fault energy. In the following we formulate a new criterion based on analysis of the elastic strain caused by lattice rotation to describe the twinnability of a hcp crystal material.

In Figure 2 we consider a pure magnesium single crystal sample with c/a ratio of 1.623. Two twin planes, the $\{10\bar{1}2\}$ plane for the π3 twin and the $\{10\bar{1}1\}$ plane for the π1 twin, lying parallel to the $[\bar{1}2\bar{1}0]$ direction, are shown in Figure 2a. It can be seen atoms in the layers along the $[\bar{1}2\bar{1}0]$ direction with the sequence of …ABAB…. Atoms in layers A and B are marked as black open and red close circles, respectively. Thus, the atomic arrangement along the $[\bar{1}2\bar{1}0]$ direction can be displayed in one plane, as shown in Figure 2b and c. Two twin planes are indicated as blue and green lines. To calculate the strain we need only analyze the crystal structures before and after twinning. We consider an elastic plane of unit thickness in a generalized plane strain state subject to a load. We select an initial representative unit $M$ shown in Figure 2b, c, which is the elementary cell of the periodic structure in the $\{\bar{1}2\bar{1}0\}$ plane. The side lengths of unit $M$ are $c$ and $\sqrt{3}\,a$, respectively, parallel and perpendicular to the c-axis direction. Because of the c/a-ratio of 1.623, unit $M$ is a rectangle, rather than a square, as shown in the left side Figure 2b and c with $L_x = 1.732a$ and $L_y = 1.623a$ (the ratio of side lengths is enlarged in this scheme to display clearly the rectangular structure).

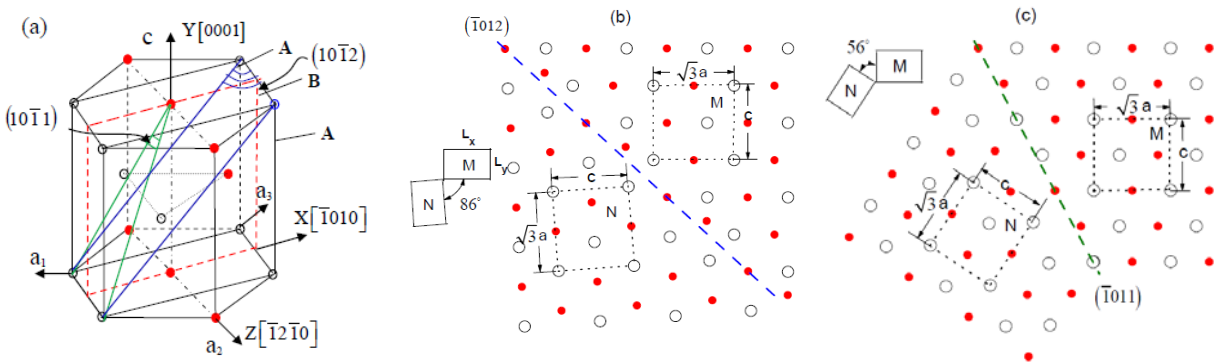

**Figure 2.** (a) Two twin planes π3 $\{10\bar{1}2\}$ (in blue), and π1 $\{10\bar{1}1\}$ (in green) in hcp crystal. Periodic arrangement of atoms along the $[\bar{1}2\bar{1}0]$ direction with the sequence of …ABAB….is shown. (b) The $\{\bar{1}012\}$ twin is symmetrical about the $(\bar{1}012)$ twin plane. (c) The $\{10\bar{1}1\}$ twin is symmetrical about the $(10\bar{1}1)$ twin plane. The representative unit $M$ is in the original matrix and *N* in the twinned region.

After the construction of this representative unit, we calculate the lattice strain $\varepsilon_c$ in the c-axis direction due to the lattice rotation after twinning (see associated content: the appendix). The result is

$$\varepsilon_c = \frac{L_y' - L_y}{L_y} = \sqrt{\frac{(L_x)^2 \sin\alpha + L_x L_y \cos\alpha}{(L_y)^2 \sin\alpha + L_x L_y \cos\alpha}} - 1, \quad (1)$$

where $L_y'$ is the equivalent unit lengths in the *y*- direction after the rotation. According to equation (1) if $L_x > L_y$ ($\alpha$=0~90°), $\varepsilon_c$ will be always positive. This means that as a result of lattice rotation the crystal will elongate along the loading direction and contract in the perpendicular direction. Under tensile loading, the lattice elongation parallel to the loading direction will facilitate the extension strain, thus favoring twinning. On the other hand, under compression loading, the lattice elongation parallel to the loading will offset the effect of twinning, and thus making twinning difficult under compression. When $\varepsilon_c$ is negative ($L_x < L_y$), the situation is in the opposite. The crystal will contract in the loading direction while elongate perpendicular to the loading direction. Under this condition, extension twinning would not be favored.

Using the new criterion, we can examine twinning in an hcp magnesium crystals, with $c/a$=1.623, for which the representative unit is a rectangle with $L_x > L_y$ as shown in Figure 2, where $L_x = \sqrt{3}a$ (1.732*a*) and $L_y = c$ (1.623*a*). The misorientation angles for the $\{10\bar{1}2\}$ twin and the $\{10\bar{1}1\}$ twin are 86.3° and 56°, respectively. Under *c*-axis loading, the lattice strain along the *c*-axis is obtained from equation (1) as $\varepsilon_c = 0.0626$ for the $\{10\bar{1}2\}$ twin, which is positive and close to the value of 0.065 according to Schmid and Boa [20]. For the $\{10\bar{1}1\}$ twin, the lattice strain along the *c*-axis direction is $\varepsilon_c = 0.0396$, which is again positive. Thus the crystal will elongate parallel to and contract perpendicular to the *c*-axis direction due to the lattice rotation for both $\{10\bar{1}2\}$ and $\{10\bar{1}1\}$ twins. When a compression is applied along the c-axis, the lattice elongation along the c-axis direction counteracts against the effect of twinning, thus making twinning more difficult. In other word, only the tensile strain along the *c*-axis direction can be accommodated by both $\{10\bar{1}2\}$ and $\{10\bar{1}1\}$ twins.

Based on our analysis, we formulate a new criterion for the possibility of twinning in hcp metals at the nanoscale. For a single crystal, the change in strain energy due to lattice rotation associated with twinning will play an important role on the plastic deformation mechanisms. If lattice elongation along the loading direction occurs due to twinning, this twin will be favored to be activated under extension. Similarly, if lattice contraction occurs along the loading direction, twinning will favored under compression. In contrast to previous analyses of strain caused by twinning shear according to the strain ellipsoid,[13,14] we propose to focus on strain-compatibility analysis of lattice rotation caused by twinning. A specific consequence is both $\{10\bar{1}2\}$ and $\{10\bar{1}1\}$ twinning in magnesium single crystal prefer to occur under c-axis tension. Our analysis is for an ideal single crystal at the nanoscale. For more complicated situations other factors may need to be considered, such as the stacking-fault energy, the free surface orientation, the loading condition, the nucleation of twinning, and so on.

Moleular dynamics (MD) simulation of deformation in hcp magnesium single crystal have been performed which gave results that may be used to test the present criterion on twinning.[11] Simulations showed that under compression loading <c+a> pyramidal slip dominates and no compression twins observed in simulations at different temperatures for different loading and boundary conditions. Besides the MD findings we can also look to the microcompression measurements of single crystal magnesium where no twinning was observed in sample sizes from 2.1 to 10 μm under c-axis compression, while significant plasticity and hardening occurred.[2,3] Moreover, both $\{10\bar{1}2\}$ tension twins and $\{10\bar{1}1\}$ compression twins have been observed in the simulation of c-axis tension [11]. Thus all the foregoing results are consistent with our theoretical analysis that shows both $\{10\bar{1}2\}$ and $\{10\bar{1}1\}$ twins can be activated under c-axis tension, and compression twins will not occur when the c/a-ratio of the hcp metal is less than $\sqrt{3}$.

To further test the validity of our criterion we present 3-D MD simulations of magnesium nanopillar under c-axis compression. Two single-crystal nanopillar models, cylindrical and square in cross section, are studied, with the diameter and side-length 15 nm and 10nm, respectively, and a length-to-diameter ratio of 2:1. The number of atoms in the two systems are 250620 and 102400 atoms, respectively. A uniaxial compression is performed along the c-axis (the long axis) by imposing compressive displacements on atoms along the c-axis with the strain rate of $1\times10^9/s$. The simulations are carried out until the maximal strain reached about 30% while the temperature is controlled at 10K by the Nosé-Hoover thermostat. Figure 3 shows the atomic configurations of magnesium nanopillar under different compressive strains. In Figure 3a, b, d, we see the first slip nucleates at the free surface initially and then travels across the pillar on the pyramidal plane. Only the leading partials and stacking faults can be found. Fig. 3c and e show the nucleated dislocations escape from the pillar, leaving clearly slip steps on the surface. Furthermore, a slight asymmetric in-plane shearing is observable, implying basal slip also plays an important role in the plastic deformation, which is consistent with recent microcompression tests.[2] Detailed investigations over the range of the current strain levels and during the entire deformation process reveal no indications of compression twins, instead indications are, the nucleation and multiplication of <c+a> pyramidal slip dominates the deformation under c-axis compression. Additionally, we have tested other hcp metals with various c/a-ratios. In Figure 4 we represent the compression deformation behaviors of titanium with c/a-ratio of 1.588. The results are similar to those on magnesium, namely, no twins could be observed under c-axis compression, and <c+a> slip along the pyramidal planes is main mechanism of plastic deformation.

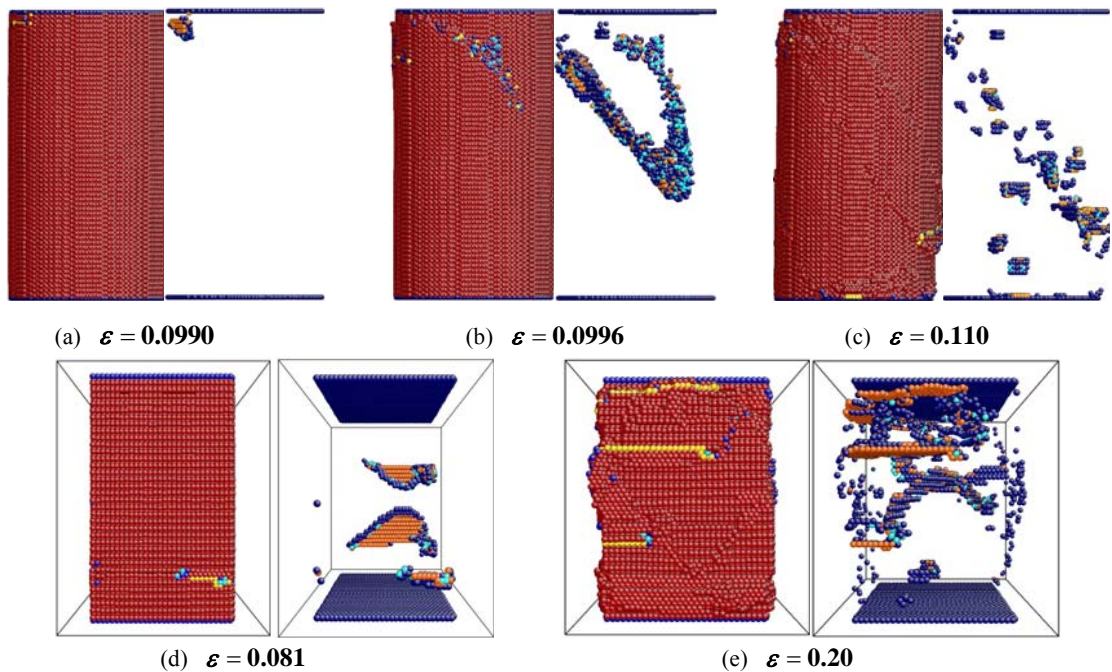

(a) $\varepsilon = 0.0990$   (b) $\varepsilon = 0.0996$   (c) $\varepsilon = 0.110$

(d) $\varepsilon = 0.081$   (e) $\varepsilon = 0.20$

**Figure 3**. (a) (b) (c) Deformation of the magnesium cylinder nanopillar under c-axis compression at different strains (D=15 nm, T=10 K). (d) (e) Deformation of the magnesium square nanopillar under c-axis compression at different strains (D=10 nm, T=10 K). An EAM potential for magnesium developed by Liu et al [22] and the molecular dynamics program LAMMPS [23] are employed. Atomeye software [24] and Common Neighbor Analysis (CNA) [25] are used to highlight the microstructure evolution. Atoms on perfect hcp lattice are shown in red (which are not shown in the right figures), atoms on fcc lattice are shown in yellow, atoms on bcc lattice are shown in cyan while others are in blue.

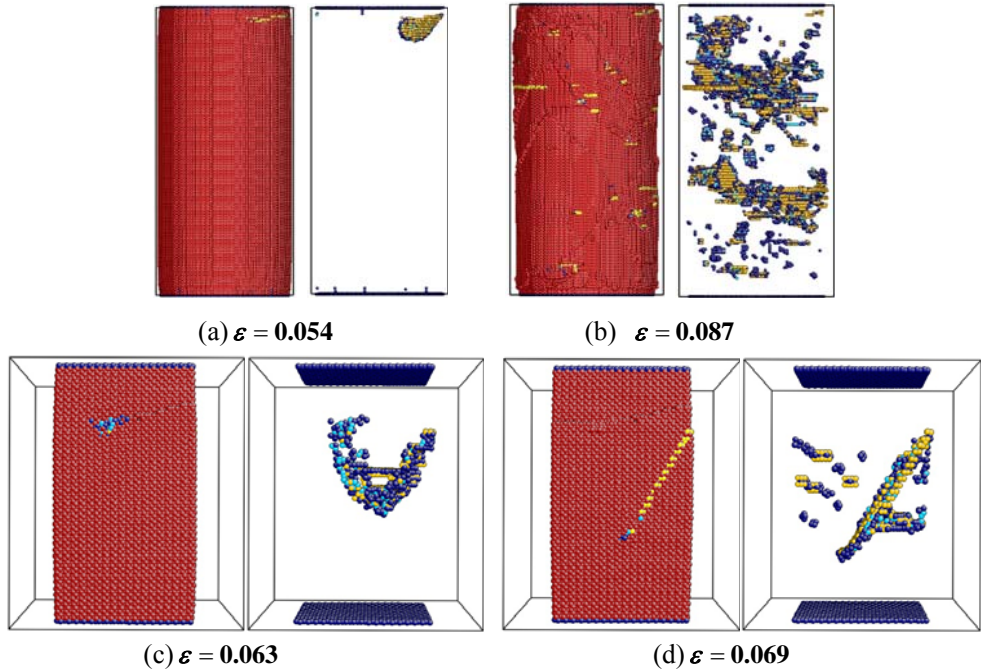

(a) $\varepsilon = 0.054$    (b) $\varepsilon = 0.087$

(c) $\varepsilon = 0.063$    (d) $\varepsilon = 0.069$

**Figure 4.** (a) (b) Deformation of the titanium cylinder nanopillar under c-axis compression at different strains (D=12 nm, T=10 K). (c) (d) Deformation of the titanium square nanopillar under c-axis compression at different strains (D=12 nm, T=10 K). An EAM potential [21] and the LAMMPS [23] are employed. The Atomeye software [24] and CNA [25] are used to highlight the microstructure evolution as stated in Fig.3.

In conclusion, our new criterion based on the analysis of the lattice rotation provides a consistent explanation for the recent experimental phenomena at the nanoscale, such as the size effect of the twinning, and the "loss" of twins. Magnesium and its wrought alloys show a pronounced direction-dependence of plastic yielding and work hardening, as well as different yielding behavior in tension and compression. The so-called strength differential (SD) effect [5] is of interest because it influences the application of magnesium and its alloys. According to our criterion, no twins should occur under the c-axis compression in a perfect magnesium single crystal at the nanoscale, while twinning dominates the deformation under the c-axis tension. This means that the confounded tension-compression asymmetry for magnesium and its alloys is closely related with the difference of deformation mechanisms, respectively, under tension and compression.

**ASSOCIATED CONTENT**
**Supporting Information**

An appendix on how to calculate the lattice-rotation strain attached. This material is available free of charge via the Internet at http://pubs.acs.org.

## AUTHOR INFORMATION
**Corresponding Author**

*E-mail: yfguo@bjtu.edu.cn.

**Author Contributions**

▫These authors contributed equally to this work

## ACKNOWLEGEMENTS

This work is supported by Chinese Nature Science Foundation (11072026) and the Fundamental Research Funds for the Central Universities of China (2013JBM009). XZT thanks the China Scholarship Council (CSC) for financial assistance and the hospitality of the Nuclear Science and Engineering Department at MIT for an academic visit.

## REFERENCES

(1) Yu, Q.; Shan, Z.W.; Li, Ju.; Huang, X.; Xiao, L.; Sun, J.; Ma, E. Strong crystal size effect on deformation twinning. *Nature* 2010, 463, 335−338.

(2) Lilleodden, E. Microcompression study of Mg (0001) single crystal. *Scripta Mater* 2010, 62, 532-535.

(3) Byer, C.M.; Li, B.; Cao, B.; Ramesh, K.T. Microcompression of single-crystal magnesium. *Scripta Mater* 2010, 62, 536-539.

(4) Syed, B.; Geng, J.; Mishra, R.K.; Kumar, K.S. [0001] Compression response at room temperature of single-crystal magnesium. *Scripta Mater* 2012, 67(7–8), 700-703.

(5) Roberts, C.S., *Magnesium and its alloys;* Wiley: New York, 1964.

(6) Avedesian, M.M.; Baker, H. *Magnesium and Magnesium Alloys, ASM Specialty Handbook;* ASM International, Materials Park: Ohio, 1999.

(7) Barnett, M.R. Twinning and the ductility of magnesium alloys Part I: "Tension" twins. *Mater Sci Eng A* 2007, 464, 1-7.

(8) Kelley, E.W.; Hosford, W.F. *Trans Metall Soc AIME* 1968, 242, 5-13.

(9) Reed-Hill, R.E.; Robertson, W.D. Additional modes of deformation twinning in magnesium. *Acta Metall* 1957, 5, 717-727.

(10) Nave, M.D.; Barnett, M.R. Microstructures and textures of pure magnesium deformed in plane-strain compression. *Scripta Mater* 2004, 51, 881-885.

(11) Guo, Y.F.; Tang, X.Z.; Wang, Y.S.; Wang, Z.D.; Yip, S. Compression Deformation Mechanisms at the Nanoscale in

Magnesium Single Crystal. *Acta Metall. Sin. (Engl. Lett.)* 2013, 26, 75-84.

(12) Guo, Y.F.; Wang, Y.S.; Qi, H.G.; Steglich, D. Atomistic simulation of tension deformation behavior in magnesium single crystal. *Acta Metall. Sin. (Engl. Lett.)* 2010, 23, 370-380.

(13) Hertzberg, R.W. *Deformation and fracture mechanics of engineering materials, fourth edition,* John Wiley & Sons, Inc., 1996.

(14) Yoo, M. H. Slip, Twinning, and Fracture in Hexagonal Close-Packed Metals. *Metall. Tran. A*, 1981, 12, 409-418.

(15) Li, B.; Ma, E. Atomic Shuffling Dominated Mechanism for Deformation Twinning in Magnesium. *Phys Rev Lett* 2009, 103, 035503-1.

(16) Wang, J.; Beyerlein, I.J.; Hirth, J.P.; Tomé, C.N. Twinning dislocations on $\{\bar{1}011\}$ and $\{\bar{1}013\}$ planes in hexagonal close-packed crystals. *Acta Mater* 2011, 59, 3990-4001.

(17) Meyers, M.A.; Benson, D.J.; Vöhringer, O.; Kad, B.K.; Xue, Q.; Fu, H.H. Constitutive description of dynamic deformation: physically-based mechanisms. *Mater. Sci. Eng. A* 2002, 322, 194-216.

(18) El-Danaf, E.; Kalidindi, S.R.; Doherty, R.D. Influence of grain size and stacking-fault energy on deformation twinning in FCC metals. *Metall. Mater. Trans. A* 1999, 30, 1223-1233.

(19) Chen, M.; Ma, E.; Hemker, K.J.; Sheng, H.; Wang, Y.; Cheng, X. Deformation twinning in nanocrystalline aluminum. *Science* 2003, 300, 1275-1277.

(20) Christian, J.W.; Mahajan, S. Deformation twinning. *Progr. Mater. Sci.* 1995, 39, 1-157.

(21) Schmid, E.; Boas, W. *Kristallplastizität;* Julius Springer: Berlin, 1935.

(22) Liu, X.Y.; Adams, J.B.; Ercolessi, F.; Moriarty, J.A. *Model. Simul. Mater. Sci. Eng.* 1996, 4, 293-303.

(23) Plimpton, S. Fast parallel algorithms for short-range molecular dynamics. *J Comput. Phys.* 1995, 117, 1-19.

(24) Li, J. AtomEye: an efficient atomistic configuration viewer. *Model. Simul. Mater. Sci. Eng.* 2003, 11, 173-177.

(25) Honeycutt, J.D.; Andersen, H.C. Molecular dynamics study of melting and freezing of small Lennard-Jones clusters. *J. Phys. Chem.* 1987, 91, 4950-4963.

**Appendix:**

**Calculation of the lattice strain caused by lattice rotation**

As we have known that the deformation twinning occurs when a region of crystal is transformed into its mirror counterpart by the external loading. As shown in App. Figure 1, the twinned region is symmetrical to the original matrix about the twin plane, while crystal rotation occurs with the misorientation angle α with no volume change after twinning. If we construct a representative unit $M$ in the untwined crystal (the elementary cell of the lattice structure), $M$ will transfers to unit $N$ after the lattice rotation. Therefore, the lattice strain along the $x$- or $y$- direction due to the lattice rotation can be calculated varied with the misorientation angle α.

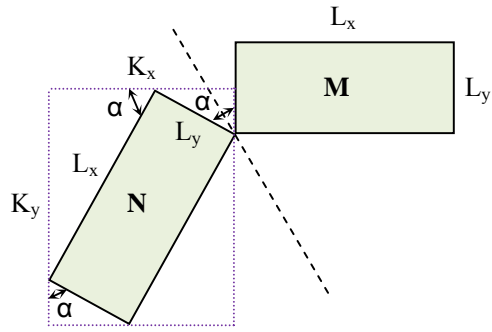

**App. Figure 1**

At first, we define $K_x$ and $K_y$ as the characteristic lengths of the characteristic unit which can tolerates the unit $N$ after the rotation as shown in App. Figure 1. The equivalent unit length after the rotation are defined as $L_x^{'}$ and $L_y^{'}$. Because there is no volume change after twinning, the equivalent unit length $L_x^{'}$ and $L_y^{'}$ should be satisfied for the condition of

$$L_x^{'} \cdot L_y^{'} = L_x \cdot L_y \ , \tag{1}$$

while the characteristic length $K_x$ and $K_y$ are proportional to the equivalent unit length $L_x^{'}$ and $L_y^{'}$ as

$$\frac{K_x}{K_y} = \frac{L_x}{L_y}. \tag{2}$$

From equations (1) and (2), we can obtain

$$\begin{cases} (L_x^{'})^2 = \dfrac{L_x \cdot L_y \cdot K_x}{K_y} \\ (L_y^{'})^2 = \dfrac{L_x \cdot L_y \cdot K_y}{K_x} \end{cases}, \tag{3}$$

and
$$\begin{cases} K_x = L_x \cos\beta + L_y \sin\beta \\ K_y = L_x \sin\beta + L_y \cos\beta \end{cases}. \tag{4}$$

Substitute equation (4) to (3), then it is given

$$\begin{cases} L_x' = \sqrt{\dfrac{L_x L_y \left(L_x \cos\alpha + L_y \sin\alpha\right)}{L_x \sin\alpha + L_y \cos\alpha}} \\ L_y' = \sqrt{\dfrac{L_x L_y \left(L_x \sin\alpha + L_y \cos\alpha\right)}{L_x \cos\alpha + L_y \sin\alpha}} \end{cases}, \tag{5}$$

so that the lattice strain in the $x$- or $y$-direction after the rotation can be expressed as

$$\begin{cases} \varepsilon_x = \dfrac{L_x' - L_x}{L_x} = \sqrt{\dfrac{(L_y)^2 \sin\alpha + L_x L_y \cos\alpha}{(L_x)^2 \sin\alpha + L_x L_y \cos\alpha}} - 1 \\ \varepsilon_y = \dfrac{L_y' - L_y}{L_y} = \sqrt{\dfrac{(L_x)^2 \sin\alpha + L_x L_y \cos\alpha}{(L_y)^2 \sin\alpha + L_x L_y \cos\alpha}} - 1 \end{cases}. \tag{6}$$

According to equation (6), it can be obtained that if $L_x > L_y$ ($\alpha$=0~90°), $\varepsilon_y$ will always be positive and $\varepsilon_x$ negative. It means that the crystal will elongate along the $y$-direction and contract along the $x$-direction after the rotation.

If the rotation angle α is 90°, the equation (6) can be simplified to

$$\begin{cases} \varepsilon_x = \dfrac{L_y - L_x}{L_x} \\ \varepsilon_y = \dfrac{L_x - L_y}{L_y} \end{cases}. \tag{7}$$

$K_x$ ($L_x'$) will equals $L_y$ and $K_y$ ($L_y'$) equals $L_x$ after the rotation of 90°, thus the lattice elongation or contraction can be clearly observed as shown App. Figure 2. If $L_x > L_y$ ($\varepsilon_y > 0$) in App. Figure 2a, the crystal will elongate parallel to the $y$-direction and contract perpendicular to the $y$-direction. While for $L_x < L_y$ ($\varepsilon_y < 0$) in App. Figure 2b, the crystal will contract parallel to the $y$-direction and elongate perpendicular to the $y$-direction.

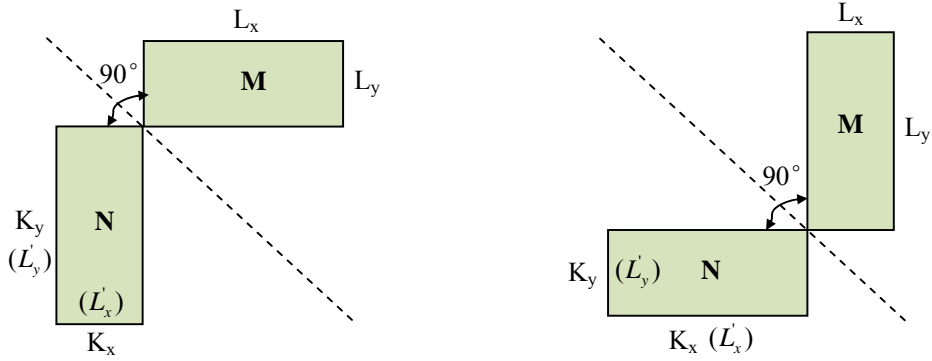

**App. Figure 2**

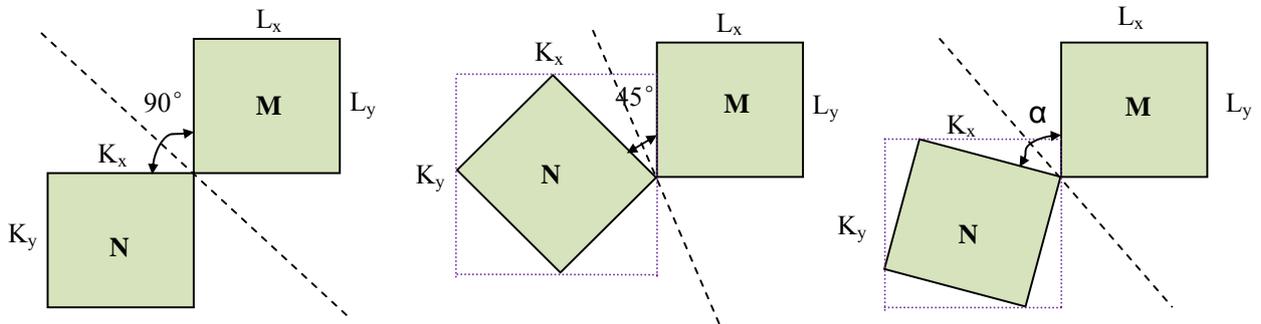

**App. Figure 3**

If the representative unit $M$ is a square ($L_x = L_y$) as shown in App. Figure 3, $\varepsilon_x$ and $\varepsilon_y$ will always equal zero according to equation (6). No elongation or contraction will occurs for the crystal along the *x*- or *y*-direction after the lattice rotation because the characteristic length $K_x$ will always equals $K_y$ with different misorientation angles α.

From above analysis, it can be concluded that after the lattice rotation with a random misorientation angle α, no lattice strain will occurs both in x- and y-directions when $L_x = L_y$. However, if the representative unit is a rectangle with $L_x \neq L_y$, the lattice strain will exists due to the lattice rotation. For $L_x > L_y$, the crystal will elongate along the *y*-direction and contract along the *x*-direction due to the rotation. While for $L_x < L_y$, the situation is in the contrary.